\documentclass[3p,times,twocolumn]{elsarticle}

\usepackage{ecrc}

\volume{00}

\firstpage{1}

\journalname{Nuclear Physics B Proceedings Supplement}

\runauth{}

\jid{nuphbp}

\jnltitlelogo{Nuclear Physics B Proceedings Supplement}

\usepackage{amssymb}
 \usepackage{amsthm}

\usepackage[figuresright]{rotating}

\usepackage{graphicx}

\begin{document}

\begin{frontmatter}



\dochead{}

\title{Gamma-rays and their future}


\author{Teresa Montaruli}

\address{Universit\'e de Gen\`eve, Facult\'e de Science, D\'epartement de Physique Nucl\'eaire et Corpusculaire}

\begin{abstract}
The focus of this presentation is to go through some of the remarkable observations concerning gamma-ray cosmic accelerators and diffuse emissions. Additionally, I will cover the status of the new generation of gamma-ray ground-based observatories, CTA and LHAASO. To create this new generation of arrays, new technologies have been prepared concerning the photosensors and the optics.
The recent start of multi-messenger astrophysical measurements indicates that there is a bright future to explore further the time domain of the universe with the current and in preparation instruments. 
\end{abstract}


\end{frontmatter}


\section{Introduction : the observation of cosmic accelerators in the gamma-ray band \label{sec:intro}}

\begin{figure}[hbt]
\centering
\includegraphics[width=0.52\textwidth]{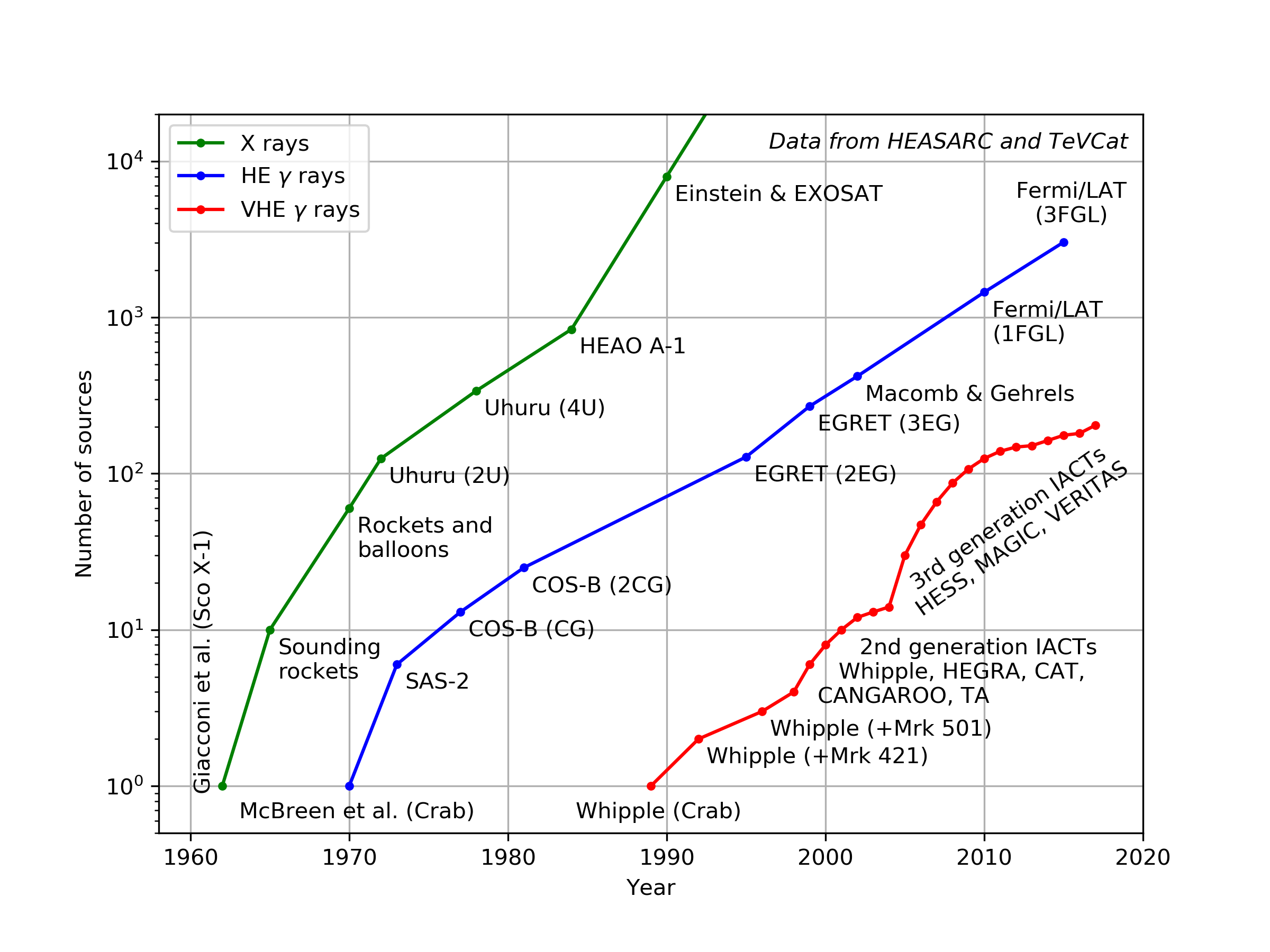}
\caption{\label{fig:kifune} X-ray and $\gamma$-ray source number versus date of detection. Plot from:  https://github.com/sfegan/kifune-plot.}
\end{figure}

\begin{figure}[hbt]
\centering
\includegraphics[width=0.5\textwidth]{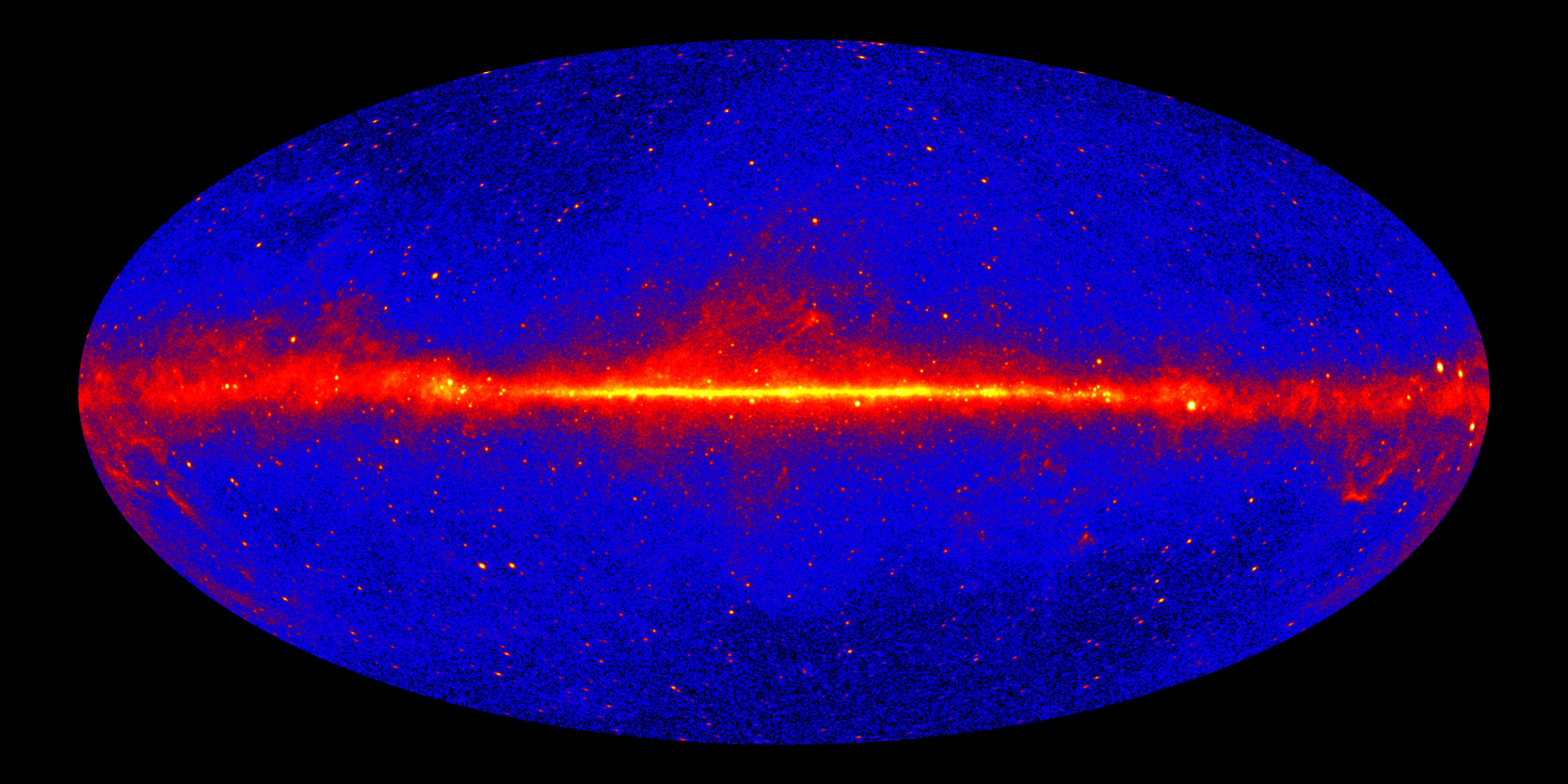}
\caption{\label{fig:fermi} The $\gamma$-ray sky of Fermi-LAT after 9 yr. Plot from:  https://svs.gsfc.nasa.gov/12969.}
\end{figure}

The number of detected high-energy gamma-ray sources is approaching that of visible stars by naked eye, according to the Kifune plot (Fig.~\ref{fig:kifune}). 
Shortly the 4FGL, now preliminarily available as FL8Y \cite{fl8y}, will contain more than 5523 sources, with emissions between energies of 100 MeV-1 TeV, detected in eight years of science data of the Fermi-LAT space detector. This has field of view (FoV) of more than 15\% of the full sky, angular resolution at 68\% containment of $\sim 0.15^\circ$ for photons with energy $> 10$ GeV and 10\% energy resolution. In normal mode Fermi would scan the full sky in 3 hours~\footnote{It was announced that, due to the failure of the motor of the solar panel in Mar. 2018, experts at NASA are  studying the best operational mode alternating periods of normal survey with periods of fixed rocking angle that allows the observation of only $\sim 1/2$ of the sky.}. This satellite mission has revolutionised our view of the high-energy sky, as it is understandable looking at the Fermi-LAT  beautiful skymap in Fig.~\ref{fig:fermi}. Of the 5523 sources observed in 8 years about 38\% has no counterpart in other wave bands and more than 52\% are active galaxies of the blazar class and 218 are pulsars. The 4FGL catalogue will superseed the published 3FGL~\cite{Acero:2015hja}, achieved with 4 years of data, containing 3034 sources with emissions beyond 100 MeV, of which 95\% are extra-galactic sources. 

\begin{figure}
\centering
\includegraphics[width=0.50\textwidth]{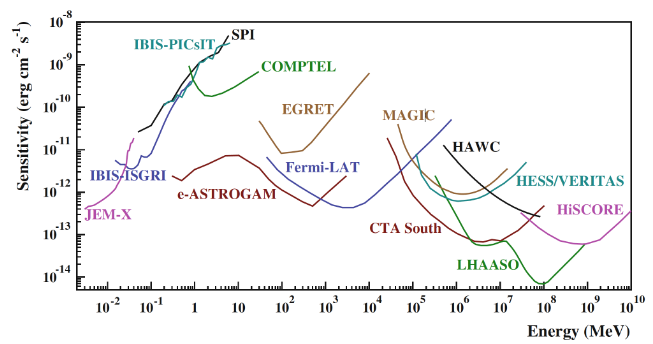}
\caption{Point source differential
sensitivity  of  different  X and  gamma-ray  instruments. For  IACTs  50 h of observation are assumed; for HAWC 5 yr and for  LHAASO
1 yr \label{fig:sens}. From Ref.~\cite{DeAngelis:2018lra}.}
\end{figure}

At higher energies ($\gtrsim 200$~GeV) the gamma-ray sky is still rich of powerful accelerators, but they cannot be observed from space due to the limited exposure of space detectors. Ground-based indirect techniques, which measure the Cherenkov light produced by the secondary products of gamma-rays, have to be adopted. Two indirect techniques of detection can be adopted:  Imaging Atmospheric Cherenkov Telescopes (IACTs), and Extensive Air shower (EAS) arrays. IACTs detect the Cherenkov light cone produced by atmospheric showers of particles when it hits the ground and EAS detect the Cherenkov light produced by the shower particles inside water tanks/ponds on ground. The two techniques are complementary, being the IACT limited in duty cycle by weather, Moon light and atmospheric conditions, while EAS are almost always sensitive. Furthermore, EAS observe about 2/3 of the sky with exposure depending on the latitude and elevation reach, while IACTs have more limited FoVs. The current generation of IACTs has FoV $\lesssim 5^\circ$ and future IACTs (with mirrors between 4-12 m diameter) will achieve $\gtrsim 8-10^\circ$. 
The main advantage of IACTs with respect to EAS is the better angular and energy resolution. The sensitivity of current and future space and ground based detectors is shown in Fig.~\ref{fig:sens}.

\begin{figure}
\centering
\includegraphics[width=0.52\textwidth]{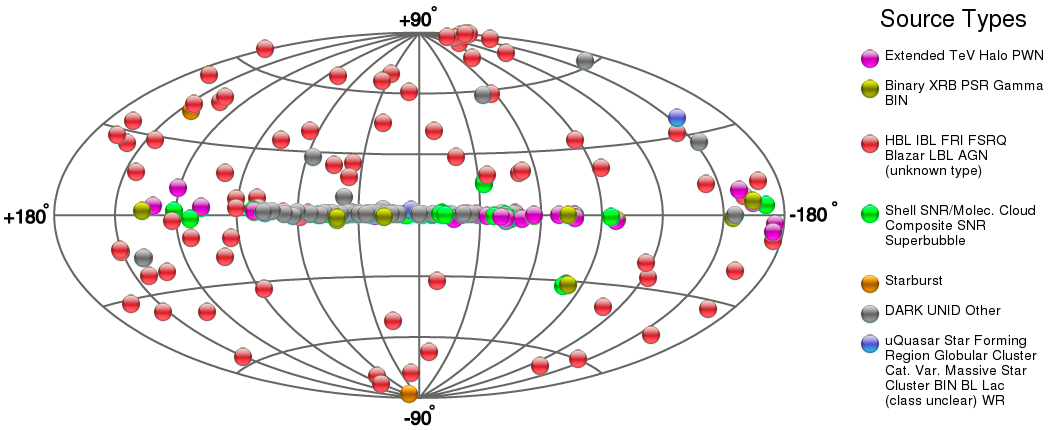}
\caption{\label{fig:tevcat} Skymap in galactic coordinates of 220 TeV cosmic sources from \cite{tevcat}.}
\end{figure}

The current most sensitive IACTs are H.E.S.S. (High Energy Stereoscopic System) \cite{HESSWeb}, MAGIC (Major Atmospheric Gamma Imaging Cherenkov Telescope) \cite{magicWeb} and VERITAS (Very Energetic Radiation Imaging Telescope Array System) \cite{VERITASWeb}, which provided most of the 220 sources listed in the TeVCat \cite{2008ICRC....3.1341W,tevcat} (see Fig.~\ref{fig:tevcat}). Of these, 30\% are not yet associated with sources in other bands. The rest belongs to different classes:  compact objects, such as black holes and neutrons stars, including PWNe with winds  emanating from them, starburst galaxies, micro-quasars and X-ray binaries.  

\begin{figure}
\centering
\includegraphics[width=0.5\textwidth]{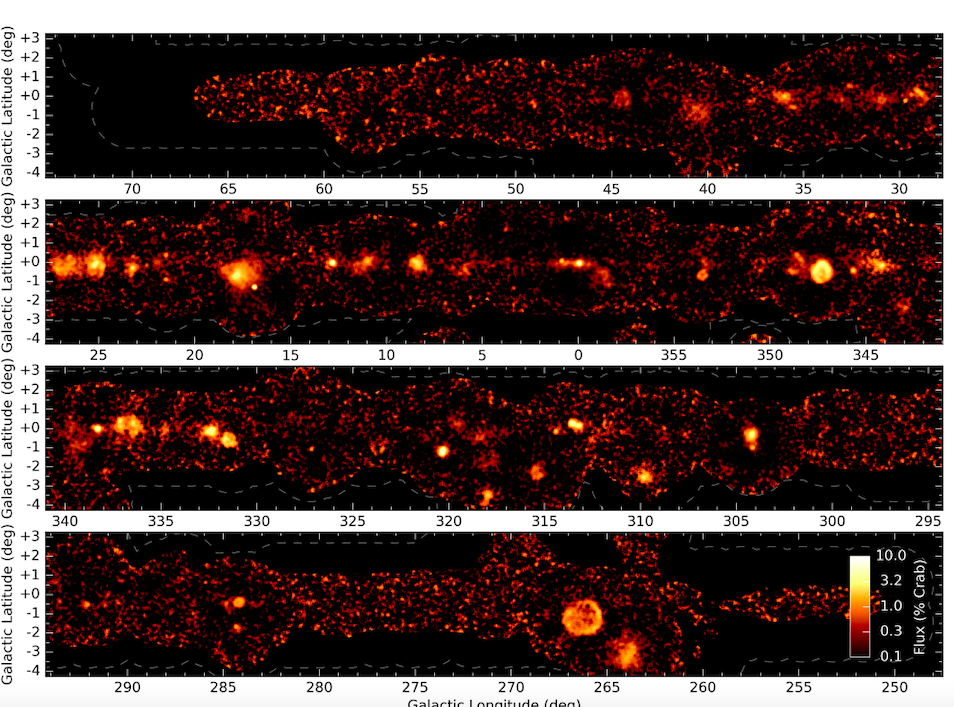}
\caption{\label{fig:hess_survey} Integral flux with energy $>1$~TeV for point sources for regions with sensitivity better than 10\% of Crab (see footnote 2) after 2700 hr of observations of H.E.S.S. \cite{2018A&A...612A...1H}.}
\end{figure}

Before the conference, in Apr. 2018. H.E.S.S. published the Galactic Plane Survey (HGPS) in Fig.~\ref{fig:hess_survey} for latitudes $|b|<3^\circ$ and longitude between $65^\circ - 250^\circ$ \cite{2018A&A...612A...1H}. This detailed map, achieving sensitivity for point-like sources in the energy range 0.2-100 TeV of $\lesssim 1.5\%$ of the Crab flux \footnote{The Crab flux above 1 TeV is a power law with  normalization $2.1 \pm 0.1_{stat} \times  10^{-11}$ cm$^{-2}$ s$^{-1}$ and spectral index $2.57 \pm 0.05_{stat}$}, was obtained with 2700 observation-hours taken from 2004 to 2013. The mean PSF at 68\% containment radius is $0.08^\circ (\sim 5$ arcmin) and the energy resolution $\sim 15\%$. While 31 sources are pulsar wind nebulae (PWNe), supernova remnants (SNRs), composite SNRs, or gamma-ray binaries, 47 sources are not yet identified but most of them (36) have possible associations with cataloged objects, notably PWNe. 

The EAS technique, as stated above, is particularly useful both to reveal extended sources/emissions and for following  flaring emissions, since it is not limited by small FoV and weather conditions. As an example, the High Altitude Water Cherenkov (HAWC) experiment \cite{HAWCWeb} and its precursor Milagro have been the sole experiments detecting gamma-rays from the Geminga pulsar. Only on Jun. 2018, at its $15^{th}$ anniversary, MAGIC finally announced the detection of Geminga pulsar pulsations by MAGIC \cite{MAGIC_Geminga}, having lowered the threshold to 30 GeV. Previously after 63 hours of observations, with threshold of 50 GeV, only limits were set \cite{2016A&A...591A.138A}. In the Geminga region two sources are identified indicating an extended region of the order of more than $2^\circ$. As a matter of fact, the test statistics is larger when an extended source of $2^\circ$ is assumed. Another similarly extended region is 2HWC J0700+143, supposed to be the 111 kyr-old pulsar PSR B0656+14. As well as Geminga, it is a close pulsar at only 288 pc distance, surrounded by the SNR Monogem ring. HAWC demonstrated that these are electron accelerators and that the produced flux is unlikely the origin of the excess of positrons measured by PAMELA and AMS-02 \cite{Abeysekara:2017old}.

The 2HWC catalogue concerns 507 d of observations contains 39 sources of which 10 are new \cite{0004-637X-843-1-40}.

\subsection{Galactic PeVatrons}

The interest in the high-energy sky is in understanding the cosmic sources capable at accelerating the CR particles that reach us with the highest energies ever observed (up to $10^{21}$~eV). These are natural laboratories to explore how the laws of physics behave at the highest energies and in extreme conditions of matter, similar to those present at the origin of the universe. The detection of the GeV-TeV radiations and cosmic particles is also key to unravel mysteries concerning the nature and location of dark matter in the universe.  
In this section, we focus on a few sources that are interesting in relation to the quest on the galactic accelerators capable of achieving PeV cosmic ray energies and so of explaining the cosmic ray knee. 
There is evidence hinting to SNRs as cosmic ray accelerators in their shock fronts, but not many between them seem capable of explaining the knee of galactic cosmic rays (CRs). Between them, RX J1713.7−3946 has the largest surface brightness, which allowed H.E.S.S to perform a comprehensive morphological study. H.E.S.S. measured the very high-energy gamma-ray spectrum in a grid of 29 small rectangular boxes of $0.18^\circ$, probing distances of the order of 0.6 pc above 2 TeV, an unprecedented accuracy in gamma-ray astronomy
\cite{Abdalla:2016vgl}. The H.E.S.S. image of RX J1713.7−3946 allows us to reveal clear morphological differences between X and gamma rays. In some regions the gamma-ray emission extends radially more than the X-ray one, hence revealing for the first time particles escaping the acceleration shock region from the outer edge of the brightest shell, a long standing prediction of Diffusive Shock Acceleration theory (for a review see Ref.~\cite{Caprioli:2015cda}). 

\begin{figure}[hbt]
\centering
\includegraphics[width=0.45\textwidth]{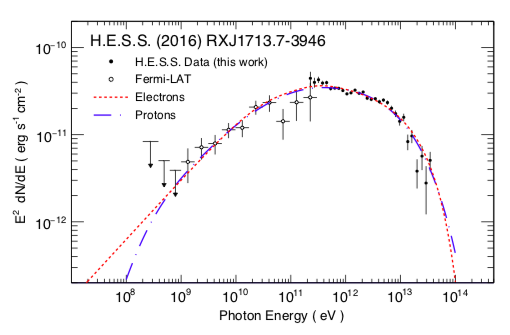}
\includegraphics[width=0.45\textwidth]{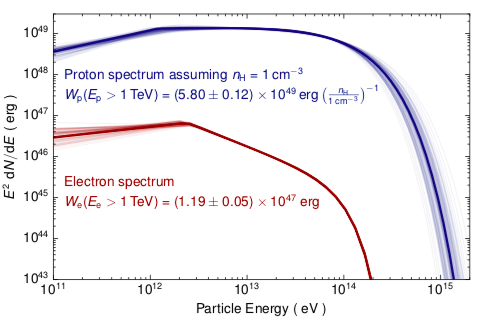}
\caption{Leptonic (red lines) and hadronic (blue lines) models  compared to Fermi-LAT and H.E.S.S. data including statistical and systematic uncertainties. The parameters of models are indicated in the bottom plot.\label{fig:rxj}  }
\end{figure}

This is new input to the discussion of the leptonic or hadronic nature of the gamma-ray emission mechanism in RX J1713.73946. In the hadronic scenario, gamma-rays come from the decay of neutral pions produced in proton-proton interactions (blue dashed line in  Fig.~\ref{fig:rxj}), while in the leptonic scenario (red dotted line),

the photons in  have up-scattered on the electron population that emits them due to synchrotron radiation. The spectrum is not conclusive in discriminating hadronic and leptonic models.

A hadronic scenario with a few TeV spectral break could be due to a  massive star exploding in a molecular cloud, itself swept away by the wind of the progenitor star, resulting in dense clumps in a lower density cavity. The energy of the break and the density of clumps depends on the SNR age. For a density of $10^3$ cm$^{-3}$ in a cavity of density 1 cm$^{-3}$ the break is between 1-5 TeV, consistent with the best fit value (see Fig.~\ref{fig:rxj}). For leptonic models, considering the SNR age of 1000 yr, a magnetic field of 70 $\mu$G seems plausible for a cooling break corresponding to the observed one, while the simultaneous fit of X-ray and $\gamma$-ray data seems to hint at lower values.
Surely, the Cherenkov Telescope Array (CTA) will provide further insights~\cite{Acero2017}. 

Very interestingly HAWC  observes four candidate galactic PeVatrons with significance well above $5\sigma$ assuming $0.5^\circ$ extension and with emission extending above 56 TeV \cite{HAWCPeVatrons}. These four regions are potential candidate galactic CR sources responsible for the knee: 

\begin{enumerate}
\item MGRO J2019+37, is a giant gamma-ray-emitting complex with 80\% of the Crab Nebula flux at 20 TeV in the Cygnus region, first observed by HAWC precursor Milagro. In the region  VERITAS resolved three TeV emitting regions coincident with Fermi sources \cite{2014ApJ...788...78A}, the harder of which is VER J2019+368 coincident PSR J2021+3651 (red region in Fig.~\ref{fig:Cygnus}). Close-by, there is a young massive stellar cluster  Sh 2-104 (blue circle in Fig.~\ref{fig:Cygnus}) from which {\it NuSTAR} observed hard X-rays \cite{2016ApJ...826...25G,2014ApJ...788...78A}. This region could host a hidden active galaxy or a pulsar. What object Sh 2-104 is and what is its interplay with the gamma-ray emission is an intriguing question, and could shed light on other such clusters found close also to other gamma-ray emitting regions (e.g. Westerlund 1 or the Carina Nebula, most probably powered by the colliding wind-binary $\eta$ Carinae). 
It is noticeable that the clusters of massive stars have been proposed as a viable solution to accelerate the galactic CRs up to the knee. Sufficient kinetic energy would be supplied by interacting stellar winds \cite{Aharonian:2018oau}.

\item MGRO J1908+06 is most probably associated to the  pulsar PSR J1907+0602 of about 20-40 kyr age, as confirmed by Fermi. It has been also observed by H.E.S.S., VERITAS and ARGO-YBJ (see Ref.~\cite{0004-637X-843-1-40} and references therein). VERITAS measured the TeV emission from PSR J1907+0602 extending toward SNR G40.5–0.5 with spectral index of -2.2 and cut-off at about 20 TeV. The lack of softening of the spectral index in J1908 far from the pulsar may argue towards the presence of another TeV source in the region \cite{2014ApJ...787..166A}.

\begin{figure}
\centering
\includegraphics[width=0.48\textwidth]{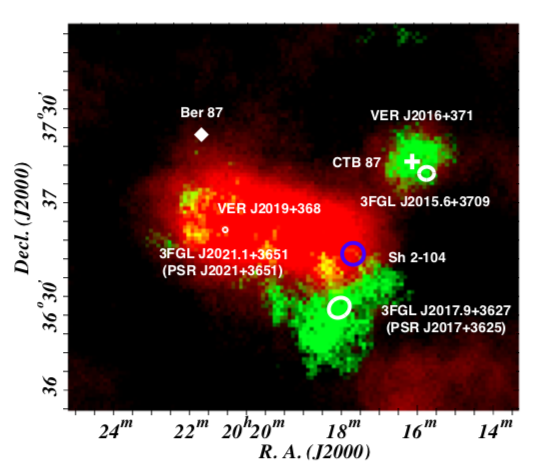}
\caption{Detailed map of MGRO J2019+37 region resolved into distinct sources by VERITAS (green region is 0.6-1 TeV emission and red is $>1$~TeV). Figure from Ref.~\protect\cite{2016ApJ...826...25G} adapted from Ref.~\protect\cite{2014ApJ...788...78A}.\label{fig:Cygnus}}
\end{figure}

\item HESS J1826-130/HESS J1825-137 are confused by HAWC. The centroid of HAWC is closer to HESS J1826-130, still an unidentified source.
It is shown in Ref.~\cite{2017arXiv170804844A} that this source, with HESS J1641-463 and HESS J1741-302, having a relatively hard spectrum extending above 10 TeV, must be capable of accelerating the parental particle population up to $>100$~TeV. Assuming a hadronic scenario, dense gas regions can provide rich target material for $pp$ interactions and consequently produce gamma-rays from $\pi^0$-decay and neutrinos from $\pi^{\pm}$-decay. Investigations on the interstellar medium along the line of sight for these sources point out the existence of dense interstellar gas structures coincident with the best fit positions of these sources. One can find possible hadronic models with CRs being accelerated close to PeV energies to explain the $\gamma$-ray emission from all of these sources. This opens up the possibility that a population of PeVatron CR accelerators might be active in the Galaxy. CTA will have a key role to find out the cut-off above 10 TeV of HESS J1826-130 and in disentangling eventual hadronic components.

\item HESS J1825-137 is an extremely bright PWN with an intrinsic diameter of ∼100 pc, potentially the largest $\gamma$-ray PWN currently known \cite{2018arXiv181012676H}. 
\end{enumerate}

The above mentioned HESS J1641-463, found within the bounds of SNR G338.5+0.1, is still an unidentified object but likely a PWN or SNR. It is among the best PeVatron candidates with a rather hard spectrum of $-(2.07 \pm 0.11_{stat} \pm 0.20_{stat})$, with 1.8\% of the Crab flux at $>1$~TeV, extending beyond 10 TeV without any sign of curvature. The H.E.S.S. collaboration has used this result to determine a 99\% C.L. cut-off lower limit of the proton spectrum at 100 TeV.

H.E.S.S. collected in 10 years high statistic measurement of the region of radius $\sim 200$ pc from the Galactic Centre with arcminute angular resolution. It tracked 
the presence of PeV particles within the central core of 10 parsec of the Galaxy. It also reconstructed a $1/r$ profile of the CR density up to 200 pc, which indicates a quasi-continuous injection of protons into the central molecular zone from an accelerator at the galactic centre on a timescale exceeding the characteristic time of diffusive escape of particles from the central molecular zone \cite{2016Natur.531..476H}.
The most probable PeVatron accelerator responsible of these particles is Sgr A$^*$. Another possible candidate arises from Chandra observations, that resolved the G359.95-0.04 PWN with a non-thermal X-ray spectrum with luminosity between 2-10 keV, about 10 times smaller than the TeV emission observed by H.E.S.S.. This has been examined as possible candidate in Ref.~\cite{2015arXiv151101159K}.
$\it NuSTAR$ detected 3.76 keV pulsations probably from a magnetar SGR J1745- 29 at $\sim 0.3 pc$ distance from Sgr A*. These are interpreted as the result of Faraday rotation from the diffuse hot gas in a 8 mG magnetic field at the 0.1 pc scale. But, corresponding to such a high magnetic field, the TeV emission resulting from the synchrotron emission normalized to the observed X-ray in the keV region, would be much lower than what observed by H.E.S.S.. Thus one would be left with needing a much weaker field within the PWN, which would be rather peculiar. As a matter of fact, for other objects in the outer galaxy it is found that the field resulting in the synchrotron radiation is at least as large as the ambient interstellar matter field. From this, the conclusion that the PeVatron is most probably the black hole Sgr A*. 

\subsection{Diffuse $\gamma$-ray emissions}

\begin{figure}
\centering
\includegraphics[width=0.50\textwidth]{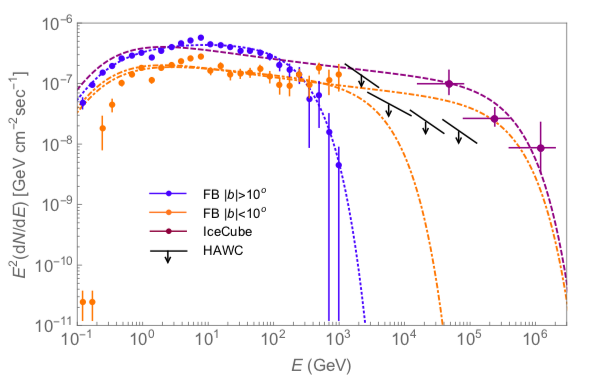}
\caption{Fermi-LAT spectra from low- and high-latitude regions of the Fermi-bubbles. Also shown are HAWC upper limits in the $b > 6^\circ$ region and neutrino spectra, assumed to originate from 8 IceCube events in the region (magenta dashed line). The sensitivity of CTA and LHAASO have been added to the plot in \cite{Yang:2018bfb} \label{fig:bubbles}}
\end{figure}

Gamma-rays are also observed from diffuse regions. They include the extended emission due to CRs propagating in the galactic plane and the mysterious Fermi Bubbles emanating up to about 25'000 light years out of the galactic centre (also visible in Fig.~\ref{fig:fermi}). 
After their discovery in the Fermi-LAT data \cite{0004-637X-717-2-825}, they were associated in various wavelengths between which the WMAP haze \cite{Finkbeiner:2003im}, confirmed by Planck \cite{Ade:2012nxf}. 
The gamma-ray power observed in the 1-100 GeV in the Bubbles is $\sim 4 \times 10^{37}$ erg s$^{-1}$, enough to cool the amount of gas that could create $2 \times 10^6 M_\odot$ by racing it through the bubble at 1000-1300 km/s \cite{2017ApJ...834..191B}. Nonetheless, this velocity value is still debated, since other measurements derive lower values \cite{2016ApJ...829....9M,2017MNRAS.467.3544S}, and this influences the date of the event (for these values around 6-9 Myr ago). 
The gamma-ray emission has a hard spectrum with $\sim -2$ spectral index and has uniform spectra up to $\pm 10^\circ$ in Galactic latitude, as visible in Fig.~\ref{fig:bubbles}, and has sharp edges.

Ref.~\cite{0004-637X-717-2-825} discusses possible models to explain the bubbles. In hadronic models, gamma-rays are produced by inelastic collisions between cosmic ray protons and thermal nuclei, via decay of neutral pions. In leptonic models, gamma-rays are generated by inverse-Compton scattering of the interstellar radiation field by CR electrons (CRe). Moreover, they might have produced by the activity of the black hole in the galactic centre or nuclear star formation. For instance, this event may have been due to a cloud of gas in-falling into the black hole between 6-9 million of years ago, causing fired off jets of matter, forming the twin lobes of hot gas seen in X-ray and gamma-ray observations \cite{2017ApJ...834..191B}. 
Cosmic rays could either be accelerated at the galactic centre or  transported to the surface of the bubbles, or accelerated in-situ by shocks or turbulence. 
Authors in \cite{0004-637X-717-2-825} point out that hadronic models fail to reproduce the observed microwave haze and require another population of primary CRe in order to match the haze emission with its soft spectrum. A giant reverse shock could be a plausible source of primary CRe that are responsible for the microwave and radio polarization data \cite{Crocker:2014fla}.

Razzaque and Yang \cite{Yang:2018bfb} normalized an hadronic model to Fermi-LAT 6.5 yr data, which assumes that the Bubbles also produce 8 high-energy starting neutrino events of IceCube  \cite{Aartsen2013}. These events happen to have directions compatible with the broad extension of the bubbles (see  Fig.~\ref{fig:bubbles} from \cite{TheFermi-LAT:2017vmf}). Indeed this assumption is quite optimistic since it ignores possible atmospheric muon and neutrino backgrounds \cite{Aartsen2013}.
Nonetheless, the resulting hadronic model is compatible to gamma-ray constraints from HAWC \cite{Abeysekara:2017wzt}. 
The synergy between the future CTA and the LHAASO observatories, one with very good sensitivity and angular resolution and the other with large FoV and high duty cycle, can be beneficial to observe the Fermi Bubbles and either detect 1 TeV emission from them or constrain models, as shown in \cite{Yang:2018bfb}. With larger FoVs, it will be possible a morphology scan as a function of the latitude, which will discriminate the hadronic/leptonic scenarios.

There exists also an extra-galactic diffuse emission, not only due to the ensemble of unresolved sources, but also to the interactions of ultra-high energy cosmic rays with the cosmic background radiation.

\subsection{The EBL and IGMF}

The measurement of the TeV blazar spectra and the knowledge of their distance allows to infer the amount of light emitted by stars throughout the evolution of the universe, given the knowledge of the cross-section of the interactions of gamma-rays on the extra-galactic background light (EBL) as a function of energy. The EBL depends on the process of galaxy formation and evolution, and is hence of cosmological interest. Current IACTs and in the future CTA have sensitivity to the optical - infrared bumps of the EBL (e.g. see Fig.~1 in Ref.~\cite{DeAngelis:2018lra}). The comparison of models extracted from gamma-ray observations is in Ref.~\cite{Korochkin:2018tll} and references therein. observing over a large spectral range up to several tens of TeV with a good spectral resolution will make it possible to find out whether the observed cut-offs in the blazar spectra are intrinsic to the sources or induced by the effect of EBL absorption.

Finally, interesting constraints can be extracted on the Intergalactic Magnetic Field (IGMF). Electron pairs ($e^−e^+$) are produced via intergalactic $\gamma-\gamma$ interactions among primary TeV and EBL photons. They can upscatter the EBL photons through inverse Compton. Hence, secondary GeV-TeV photons accompanying primary TeV emissions of blazars can be detected.
Depending on the IGMF value, such secondary components may be observable either as “pair echos” that arrive with a time delay relative to the primary emission \cite{1995Natur.374..430P}, or as extended emission around the primary source  Ref.~\cite{Aharonian:1993vz,Neronov:2006hc}. A weak IGMF would not deflect much the electron pairs contrary to a strong one, which would produce GeV isotropically distributed photons around the blazar, blurring its image. Current limits set by Fermi are at the level of  $B<10^{−19}$ G for coherence length $L_B \ge 1$ Mpc at $>5\sigma$ for most EBL models \cite{Armstrong:2013hha}.
CTA turns out to be an ideal instrument to probe the nature of the IGMF and could solve the problem of the origin of magnetic fields in galaxies and galaxy clusters which is one of the long-standing unsolved problems of astrophysics and cosmology
\cite{2013APh....43..215S}.

\section{Current and future gamma-ray observatories and new techniques}

C. Galbraith and J. Jelley, when visiting the Harwell Air Shower Array in UK in 1952, used a 5 cm-diameter photomultiplier (PMT) mounted on the focal plane of a 25 cm parabolic mirror in a garbage can. They observed oscilloscope triggers from light pulses that exceeded the average night-sky background (NSB) every 2 min \cite{1953Natur.171..349G}. In 1953, from the polarisation and spectral distribution, they confirmed P. Backett’s assertion that 1 part in 10'000 of the NSB is produced by light emitted by charged CRs. After Morrison's theoretical prediction of strong gamma-ray line fluxes from gamma-ray sources in 1958, the year after G. Cocconi proposed to measure TeV gamma-rays using air shower detectors \cite{1960ICRC....2..309C}. Gamma-ray astronomy effectively started with the detection in 1989 of the Crab Nebula~\cite{1989ApJ...342..379W} by the Whipple telescope. 

The current generation of IACT and EAS arrays sees as main actors:
\begin{itemize}
\item H.E.S.S.~\cite{HESSWeb}, located in Namibia at 1800 m asl, is composed of 4 Davies-Cotton telescopes of 12 m diameter and a camera of 960 PMTs, each seeing $0.16^\circ$ of the sky. They take data since 2004. H.E.S.S. telescopes have FoV of $5^\circ$ in diameter, fully uniform to $2^\circ$ and dropping by $40\%$ of the peak value at $4^\circ$. The additional larger telescope, with equivalent mirror diameter of 28 m, saw the first light on July 2012 \cite{HESSfirstlight}.

\item VERITAS \cite{VERITASWeb}, the results of which were presented by B. Humensky at this conference, is an array of 4 - 12 m-diameter telescopes, separated by about 100 m, in Arizona at only 1250 m of altitude close to the Whipple higher site (see Fig.\ref{fig:VERITAS}). The cameras have FoV of $3.5^\circ$, achieved with 500 pixels with PMTs. The electronics is based of very fast FADCs of 500 MHz. It is taking data since more than 10 years.
\begin{figure}[hbt]
\centering
\includegraphics[width=0.50\textwidth]{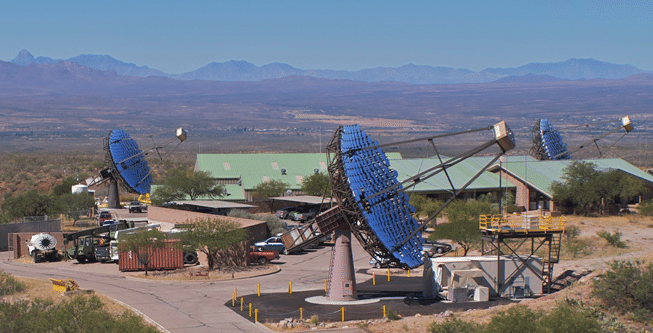}
\caption{The VERITAS array of 4 telescopes. At the site now also the CTA SCT is located (see the text). \label{fig:VERITAS}}
\end{figure}

\item MAGIC \cite{magicWeb}, the results of which are presented at this conference by M. Doro, is composed by two 17 m-diameter telescopes, located at the Observatory Los Roques de Muchachos in the La Palma island at 2200 m. MAGIC-I is operative since more than 14 yr while MAGIC-II since more than 10 years. The camera of MAGIC-II, with FoV $= 3.5^\circ$, has 1039 pixels with photomultipliers seeing $0.1^\circ$ of the sky and read by 500 MHz FADCs. The telescopes are visible in Fig.~\ref{fig:LST-1}, where also - on the left - the CTA first large size telescope (LST) and FACT \cite{FACT} are visible. 

\begin{figure}[hbt]
\centering
\includegraphics[width=0.50\textwidth]{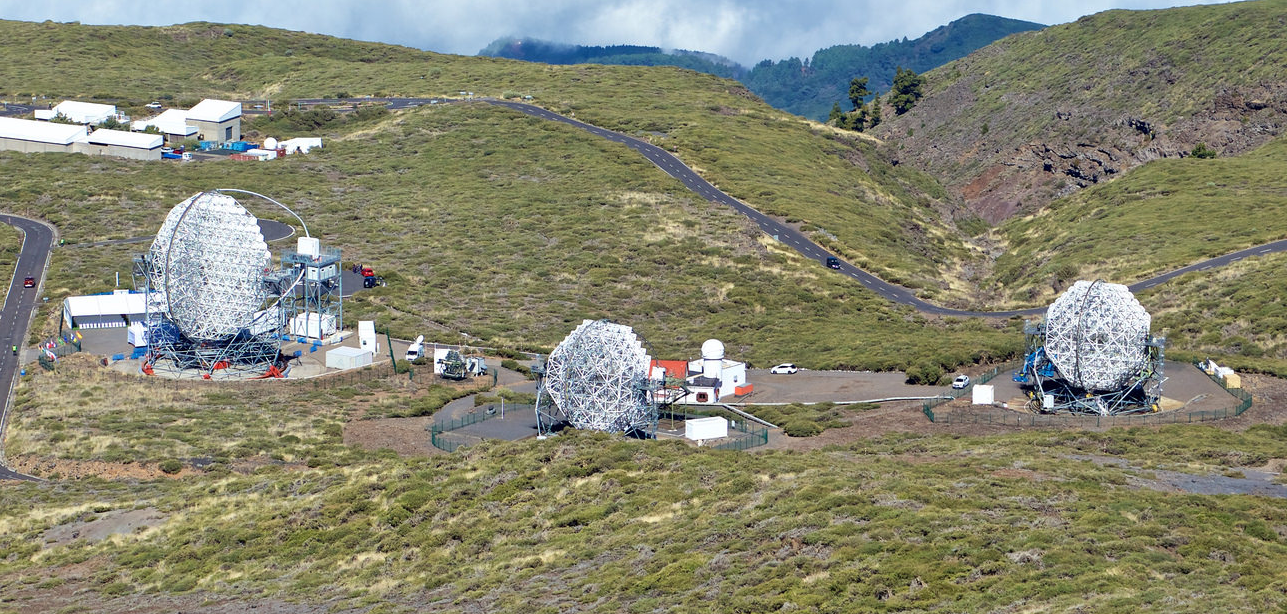}
\caption{From left to right,  the  first LST of CTA of 24~m  mirror-diameter, FACT (4  m) and  the two MAGIC telescopes (12 m) at the Observatory in La Palma.  The LST-1 was inaugurated on Oct.~2018.  \label{fig:LST-1}}
\end{figure}

\item HAWC (see Fig.~\ref{fig:HAWC}) is located close to Puebla in Mexico at an altitude of 4100 m a.s.l, with 22'000 m$^2$ collecting area and about 300 tanks, each with 4 PMTs at the bottom. They contain a total of about 20'000 litres of water. HAWC has currently best sensitivity above 10~TeV (see Fig.~\ref{fig:sens}). 
\end{itemize}
\begin{figure}[hbt]
\centering
\includegraphics[width=0.45\textwidth]{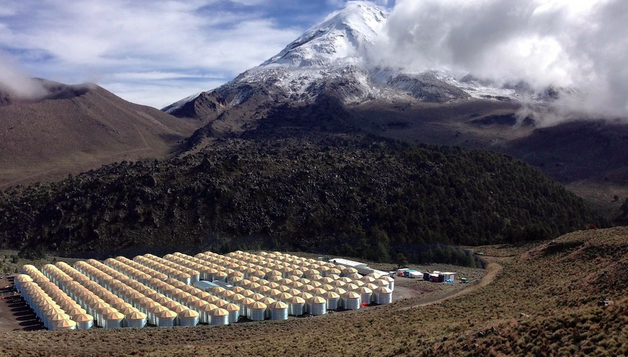}
\caption{The HAWC EAS array \cite{HAWCWeb} and in the background the volcan Pico de Orizaba.}
\label{fig:HAWC}
\end{figure}

The next generation of gamma-ray infrastructures on ground will be the CTA  \cite{CTAWeb,Acharya:2017ttl} and the LHAASO Observatories \cite{DiSciascio:2016rgi,LHAASOWeb}. 
CTAO will become an open-access observatory. Some reserved time is foreseen to ensure stable operation of telescopes and there will be a fraction of time reserved for the science key cases to the CTA Consortium described in \cite{Acharya:2017ttl}. The rest of the time will be allocated through a competitive process.

CTA will be composed of a Southern array in Chile, at the ESO site of Paranal and a Northern array in La Palma, Canarian Islands, both at about 2000 m a.s.l.. These arrays will be composed by a total of more than 100 IACTs of different sizes. The three sizes of the telescope mirrors, 24 m diameter for LSTs, 12 m for Middle Size Telescopes (MSTs) and 4 m for the small size telescopes (SSTs), ensure a wide sensitivity in energy spanning from 20 GeV to 300 TeV. CTA will start production of telescopes in 2019-20 and their installation is already ongoing in the Northern site (see Fig.~\ref{fig:LST-1} and in the Southern site it should happen in 2021. 

\begin{figure}[hbt]
\centering
\includegraphics[width=0.45\textwidth]{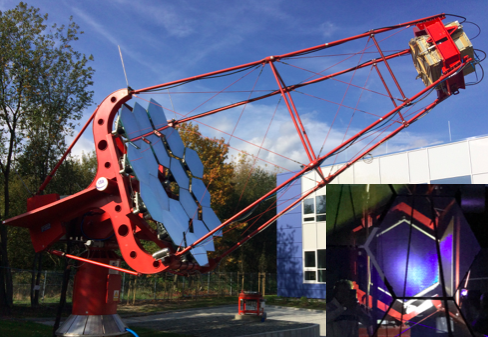}
\caption{The SST-1M prototype in Krakow and its SiPM camera in the insert.}
\label{fig:SST-1M}
\end{figure}

The Southern array will comprise a network of 70 SSTs, separated by about 200 m, covering a km$^2$ surface, to reach better sensitivity above 3 TeV. Remarkable R\&D was devoted to the SSTs, since for the first time it will be necessary to produce and operate such a large number of telescopes. 
Three SST designs have been proposed based on two possible configurations of the mirrors: a dual mirror Schwarzschild-Couder (SC) configuration, such as the one adopted by ASTRI (see Fig.~\ref{fig:ASTRI}) and by the GCT project, and a single mirror Davies-Cotton (DC) one, adopted by the SST-1M (see Fig.~\ref{fig:SST-1M}) and similar in concept to the MSTs. The SC design is aimed at a stable PSF as a function of the off-axis angle, while the DC design has degrading PSF with off-axis angle.  For a more detailed description of the SST projects see Ref.~\cite{Montaruli_icrc}.

All SSTs and the Schwarzschild-Couder Telescope (SCT)~\footnote{The SCT is a middle size telescope proposed for CTA. It is currently installed at the VERITAS site. An interesting video on the assembly of its structure is in~\cite{SCTvideo}.}  adopted a new technology as sensors of the camera, already pioneered by FACT \cite{FACT}: silicon photomultipliers (SiPMs). Time will more clearly tell the advantages of this technology, but from the prototype experience they show remarkable robustness and interesting capabilities of operation in the presence of large NSB, up to some GHz per pixel. Nonetheless, they require also appropriate calibration, as discussed in Ref.~\cite{Heller2017,Neise:2017ldg}. The duty cycle achievable with SiPM cameras seems to be larger than the duty cycle achieved by photomultiplier cameras. Photomultiplier cameras can operate in the presence of high Moon by adopting filters or by reducing the voltage applied to the PMTs and (e.g. see for MAGIC \cite{Guberman:2015rsa} and for VERITAS \cite{Archambault:2017jgj}). Since Sep. 2012, VERITAS has shown ability to observe with NSB about 30 times higher NSB levels than before increasing the observation time by about 30\%). Nonetheless, filters require human intervention. FACT recently showed that useful data can be taken with higher trigger threshold with NSB about $\sim 30\%$ above MAGIC definition of dark night. SiPM do not require human intervention on site for this, unlike it has to be done when filters are applied \cite{Neise:2017ldg}. The challenge of SiPMs is that the value of the bias sensor at the input of the sensors should be chosen based on the expected light intensity and operational voltage, and kept smaller than currently recommended in producer's manuals. As a matter of fact, this reduces the effect of the voltage drop which causes changes in operation characteristics (on photodetection efficiency, gain, crosstalk,...), while preventing overheating of the sensor.

LHAASO is under construction at 4400 m a.s.l, at Daochen where the atmosphere has 600 g/cm$^2$ depth, and should be completed in the next two years. LHAASO will be dedicated to CR measurements above the knee and its core to gamma-ray astronomy. It will see 1/7 of the sky at any moment and about 60\% of the sky every day. It will be a composite array composed by a Wide Field Cherenkov Telescope Array (WFCTA), an ensemble of 18 Cherenkov and fluorescence telescopes with cameras with 1024 SiPM based pixels; a sampling array (KM2A) of 5195 scintillator stations of 1 m$^2$, 15 m spacing, and 1171 muon counters made of 36 m$^2$ of water tanks at 30 m distance one from the other; a Water Cherenkov Detector Array (WCDA), a pond of 78'000 m$^2$ with 3125 cells of 25 m$^2$ equipped with photomultipliers of 20 inch area. These have been designed and produced in China for JUNO and LHAASO.
LHAASO will by pass the sensitivity of HAWC and also extend it to lower energy thanks to the adoption of these new large area PMTs (see \cite{DeAngelis:2018lra}). Its construction is ongoing and the pond is being filled in Jan. 2019.

\section{The Multi-messenger sources}

\begin{figure}
  \begin{center}
    \includegraphics[width=0.49\textwidth]{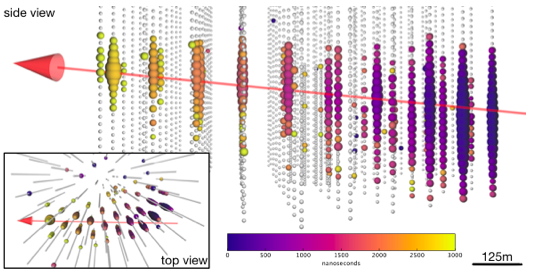}
    \caption{\label{fig:icecube} Side and top view of the multi-PeV IceCube event IC-170922A \cite{IceCube:2018dnn}. The size of the spheres is proportional to the amount of light detected by photomultipliers and  the color code indicates the photons arrival times. The arrow shows the best fit track direction. }
  \end{center}
  \end{figure}

At the time of the conference, the Science publications on the evidence for a CR source was under embargo. Nonetheless, it is worth mentioning here this important milestone in the understanding of the very-high energy CR sources~\cite{IceCube:2018dnn}.
On  September 22, 2017, IceCube  sent  an  alert  to  GCN \cite{GCN21916}. The neutrino which prompted the alert (IC-170922A) induced an up-going through-going muon track-like event. Its most probable energy was inferred to be 290 TeV with an upper limit at 90\% CL of 4.5 PeV for a spectrum of $E^{-2.13}$. This results from the fit between  194 TeV and 7.8 PeV of the cosmic neutrino signal on top of the atmospheric neutrino background in the up-going muon neutrino diffuse analysis in Ref.~\cite{IceCube:2018dnn}. Adopting this spectral shape, the probability that IC-170922A is of astrophysical origin rather than atmospheric is 56.5\%. The angular error at which the track was reconstructed was about 15 arcmin at 50\% containment.
Fig.~\ref{fig:icecube} shows the visualization (event display) of the IceCube event. 

Following  the  alert,  Fermi-LAT detected  increased gamma-ray flux from a known blazar, TXS 0506+056, located inside the directional uncertainty contour of IC-170922A \cite{IceCube:2018dnn,IceCube:2018cha}. MAGIC followed up TXS 0506+056 detecting gamma-rays with energies between $\sim 100-400$~GeV at $6\sigma$ confidence level in 12~hr of observations between Sep. 24 and Oct. 4, 2017 \cite{Ahnen:2018mvi}. The presence of a flaring source was also confirmed by VERITAS \cite{Abeysekara:2018oub}. These were the first detections of gamma rays at those energies from TXS 0506+056. Correlation between the gamma-ray emission and the high-energy neutrino is preferred over a chance coincidence at 3$\sigma$ confidence level~\cite{IceCube:2018dnn}. The redshift of the blazar was unknown prior to these observations and was measured to be $0.3365 \pm 0.0010$~\cite{Paiano:2018qeq}.
This event was reported by newspapers worldwide .

Looking in the data of a previous search for time-dependent emission all over the sky, another excess of events in the region of IC-170922A was observed, which turned our to be a longer flare between Sep. 2014 and Mar. 2015 \cite{IceCube:2018cha}. The best fit Gaussian
time window of the excess is centred on December 13, 2014, with a duration of $110^{+35}_{-24}$ days. The observed excess is $13 \pm 5$ events above the expected background from atmospheric neutrinos. The excess is inconsistent with the background-only hypothesis at the $3.5\sigma$ level.

These observations add up to a long-term observation program of IceCube concerning flaring blazars \cite{Asen2015} and stacking searches of excesses of neutrinos from blazars \cite{Aartsen:2016lir,Huber:2017wxt}. Using the above mentioned $E^{-2.13}$ power law for the 127 GeV emitting blazars in the 2FHL HBL catalogue of Fermi-LAT, the maximum contribution from that population to the IceCube observed astrophysical flux is $\sim 5\%$ at 90\% C.L.. On the other hand, when assuming that the neutrino emission strength of the individual blazars is directly proportional to their gamma-ray flux, the 2FHL HBLs can only explain 3.8\% of the diffuse $\nu_{\mu}$ flux observed by IceCube \cite{Huber:2017wxt} at 90\% C.L.. These limits can be evaded under certain assumptions, e.g. different spectra, due for instance to acceleration processes dominated by magnetic reconnection or happening in the vacuum gaps of BH magnetospheres \cite{Neronov:2016ksj}, or if it exists a peculiar class of AGNs, similar in properties to TXS 0506+056, that constitutes only 5\% of the Fermi observed blazars \cite{Halzen:2018iak}.

This example points out the importance of connecting different messengers, gamma-rays, neutrinos and also gravitational waves, to explore further the time domain of the universe and understand the highest energy sources of the universe. Some programs are already active, for instance a ToO program is in place between IceCube and MAGIC \cite{Satalecka:2017cft} and many MoUs and agreements are already in place, while waiting for the astroparticle community to get ready for the open data era of multi-messenger high-energy astrophysics.

\section{Conclusions}
There is a bright future ahead of multi-messenger high-energy astrophysics which will require a strong theory support and the adoption of open data policies to make the field flourish of interested scientists to exploit the coming large facilities that will be operetional already in the 20'ies.

My thoughts when preparing the presentation and this script went to a great and generous colleague and friend, E. Lorentz, whose work was fundamental for HEGRA, MAGIC, CTA and in general for gamma-ray astronomy.






\bibliographystyle{elsarticle-num}
\bibliography{multimessenger} 





\end{document}